# Effect of the growth temperature and the AlN mole fraction on In incorporation and properties of quaternary III-nitride layers grown by molecular beam epitaxy

S. Fernández-Garrido[1,*], A. Redondo-Cubero[1,2], R. Gago[2,3], F. Bertram[4], J. Christen[4], E. Luna[5], A. Trampert[5], J. Pereiro[1], E. Muñoz[1], and E. Calleja[1]

[1]ISOM and Dpt. de Ingeniería Electrónica, ETSI Telecomunicación, Universidad Politécnica de Madrid, Ciudad Universitaria s/n, 28040 Madrid Spain

[2]Centro de Micro-Análisis de Materiales, Universidad Autónoma de Madrid, 28049 Cantoblanco, Madrid Spain

[3] Instituto de Ciencia de Materiales de Madrid, Consejo Superior de Investigaciones Científicas, E-28049 Madrid, Spain

[4]Institute of Experimental Physics, Otto-von-Guericke-University Magdeburg, Germany

[5]Paul-Drude-Institut für Festkörperelektronik, Berlin, Germany

Indium incorporation into wurtzite (0001)-oriented $In_xAl_yGa_{1-x-y}N$ layers grown by plasma-assisted molecular beam epitaxy was studied as a function of the growth temperature (565-635 ºC) and the AlN mole fraction (0.01 < y < 0.27). The layer stoichiometry was determined by Rutherford backscattering spectrometry (RBS). RBS shows that Indium incorporation decreased continuously with increasing growth temperature due to thermally enhanced dissociation of In-N bonds and for increasing AlN mole fractions. High resolution x-ray diffraction and transmission electron microscopy (TEM) measurements did not show evidence of phase separation. The mosaicity of the quaternary layers was found to be mainly determined by the growth temperature and independent on alloy composition within the range studied. However, depending on the AlN mole fraction, nanometer-sized composition fluctuations were detected by TEM. Photoluminescence spectra showed a single broad emission at room temperature, with energy and bandwidth S- and W-shaped temperature dependences typical of exciton localization by alloy inhomogeneities. Cathodo-luminescence measurements demonstrated that the alloy inhomogeneities, responsible of exciton localization, occur on a lateral length scale below 150 nm, which is corroborated by TEM.

* electronic email: sfernandez@die.upm.es

# I. INTRODUCTION

Quaternary III-nitrides (QNs) are quite promising materials as active layers in visible and UV-light emitting diodes (LED) and laser diodes (LD) because they provide a separate control of the lattice parameter and the band gap energy. This fact enables to reduce strain-related defects and the built-in electric field present in quantum well heterostructures grown along the *c*-axis [1,2]. In addition, the introduction of a small amount of In into AlGaN has been reported to notably increase the internal quantum efficiency of UV-LEDs due to a strong exciton localization induced by In which prevents carriers to reach non-radiative recombination centers (NRCs) [3-5].

Radio frequency plasma-assisted molecular beam epitaxy (PA-MBE) offers several potential advantages over other epitaxial growth techniques including *in situ* growth control by reflection high-energy electron diffraction (RHEED), low impurity incorporation, atomically sharp interfaces, and p-type doping without need of activation process. However, in spite of these advantages, not many works have addressed the growth of QNs layers by PA-MBE [6-11]. Thus, the quality of QN layers and related devices grown by PA-MBE remains behind those grown by other epitaxial techniques, like metalorganic vapour phase epitaxy (MOVPE) [4,12,13].

This work reports on the growth and properties of QN thick layers (~ 300 nm) produced by PA-MBE with InN mole fractions below 15% and AlN mole fractions in the 1 % to 27 % range. Indium incorporation was analysed as a function on the growth temperature, between 565 and 635 ºC. The dependence of In incorporation on the AlN mole fraction was also investigated. The structural properties of the QN layers were analysed as a function of the growth temperature and the AlN mole fraction by high resolution x-ray diffraction (HRXRD) and transmission electron microscopy (TEM).

The optical properties were investigated by photoluminescence (PL) and cathodo-luminescence (CL).

## II. EXPERIMENTAL

Wurtzite (0001)-oriented QN layers were grown by PA-MBE in a Riber Compact 21 MBE system equipped with an Addon rf-plasma N source and Knudsen effusion cells for In, Ga and Al. The substrate temperature was measured using an Ircon Modline 3 optical pyrometer. As substrates (0001)-GaN templates (3.6 μm thick) grown by MOVPE on $c$-plane sapphire with a threading dislocation density in the 1-10 x $10^8$ $cm^{-2}$ range (Lumilog) were used. Their backsides were coated with Ti for efficient heat absorption during growth. After a chemical degrease, the substrates were introduced into the MBE system and outgassed during 1 hour at 400 ºC. Prior to the QN growth, a 100 nm thick GaN buffer layer was grown at 730 ºC under intermediate Ga-rich conditions [14] to obtain a smooth and flat surface. During growth the $N_2$ partial pressure in the growth chamber was kept to 1.1 x $10^{-5}$ Torr.

Alloy compositions and layer thicknesses were determined by Rutherford backscattering spectrometry (RBS) using a 2 MeV $He^+$ ion beam (~1 $mm^2$ beam spot) [15]. Combined channeling measurements (RBS/C) along <0001> axis of the wurtzite structure were carried out to determine crystal quality with depth resolution. Both random and <0001> aligned spectra were acquired in an IBM geometry with the detector placed at a scattering angle of 165º. Sample position was controlled by a 3-axis goniometer with 0.01º of accuracy. Experimental data were simulated using the RBX program [16]. Complementary characterization of the crystal quality was assessed by HRXRD and TEM measurements. TEM was carried out in a JEOL JEM 3010 microscope operating at 300 kV equipped with a GATAN CCD camera. Cross-sectional TEM specimens were prepared using standard mechanical thinning followed by Ar-ion milling.

PL was excited at 244 nm (@ 20 mW) from the 488 nm line of an Ar laser with a second harmonic generator, dispersed by a Jobin-Yvon monochromator, and detected by a UV-enhanced GaAs photomultiplier. CL measurements were carried out in a home made system. Here, the luminescence is collected by an optical grade elliptical mirror into a 50 cm grating imaging monochromator and detected using an intensified diode array.

## III. RESULTS AND DISCUSSION

### A. Indium incorporation during QN PA-MBE growth

As in the case of most of III-nitrides, in order to obtain high quality QN layers by PA-MBE it is necessary to carry out the growth under intermediate metal-rich growth conditions [9]. This growth regime is characterized by the formation of a surfactant metal adlayer with coverage values ranging from fractions of 1 ML to a 2.5-ML-thick depending on the metal excess, growth temperature and surface composition [14,17-21]. Thus, for a given growth temperature, metal fluxes must be adjusted to ensure intermediate metal-rich conditions. The growth temperature is the other critical parameter which influences adatoms mobilities and their sticking coefficients. Because III-nitrides are thermodynamically unstable at the typical pressure for PA-MBE growth, the growth temperature is limited by the decomposition rate of the growing layer [22,23]. Hence, to grow InN, GaN or AlN layers, temperatures below 500, 750 and 850 ºC, respectively, are normally used in order to keep the decomposition rate below the growth one (limited by the available impinging fluxes). In the case of QN layers, a trade-off must be considered between the high temperature desirable to grow GaN and AlN (enhaced Ga and Al mobilities), and the low temperatures necessary to avoid In-N bonds dissociation and the total re-evaporation of the impinging In flux. Typically the growth temperatures reported in the literature lie in the 550-650 ºC range [8-11]. For such a range of temperatures Al and Ga desorption rates are negligible and their sticking coefficient is unity [8]. In contrast, In atoms tend to segregate on the growing surface and its sticking coefficient is much lower than unity. Thus, during QN growth the metal adlayer is mainly provided by the In flux and the sum of Ga and Al fluxes must be kept below the impinging active N flux to enable In incorporation [8,9]. In the following, we analyse separately the effects of the growth temperature and the AlN mole fraction on In incorporation.

## 1. Effect of the growth temperature

A first series of QN layers was grown to investigate the dependence of In incorporation on the growth temperature between 565 and 635 ºC (series A). To promote a 2 dimensional (2D) growth mode, while avoiding the formation of metal droplets on the surface, all samples were grown within the intermediate metal-rich growth regime using a III/V ratio close to one. For a given active N flux, $\Phi_N$, the sum of all metal fluxes, $\Phi_{Ga} + \Phi_{Al} + \Phi_{In}$, was kept constant. The Al flux was ~ 10% of $\Phi_N$ and the sum of the Ga and Al fluxes was below $\Phi_N$ to enable In incorporation, $(\Phi_{Ga} + \Phi_{Al}) / \Phi_N \sim 75$ %. In all cases, a streaky RHEED pattern, indicative of a 2D growth mode, was observed during growth. Upon growth termination, no metal droplets were detected by optical microscopy.

Alloy compositions and thicknesses of the QN layers were determined by RBS. Figure 1 shows the measured and simulated RBS spectra for random and <0001> aligned positions in a QN sample grown at 615 ºC. The surface channels were marked in the graph, being heavier elements located at the sample surface detected at higher energies. Note that, the low signal from light elements, such as Al and N (not shown in this energy scale for clarity) in respect to the background, limits the sensitivity, arising from the higher cross-section of heavier elements such as In and Ga. For this reason, the composition was derived assuming an $In_xAl_yGa_{1-x-y}N/GaN$ structure, hence, the Al content was calculated in terms of In and Ga deficiency of a stoichiometric nitride. For growth temperatures below 615 ºC, data were fitted accurately assuming an homogeneous In incorporation along the growth direction. For higher temperatures, even with the previous assumption, an In-related peak was observed on the QN surface (inset Fig. 1). This In-related peak was simulated as a segregated In film of a very low areal density ($\sim 0.6 \cdot 10^{15}$ at/cm$^2$). This analysis is in good agreement with the results reported by Monroy *et. al.* in Ref. [9]. In addition, the In layer at the surface did not show any crystalline behavior in the RBS/C measurements along <0001> axis, what suggests a polycrystalline or an amorphous layer. Figure 2 shows the temperature dependences of both In incorporation and growth rate. The growth rates are given in units of ML/s where one ML corresponds to *c*/2. As shown in Fig. 2(a), In incorporation lay below the nominal value given by, ($\Phi_N$ - $\Phi_{Ga}$ - $\Phi_{Al}$) / $\Phi_N$ $\sim$ 25 %, and decreased steadily with the growth temperature from $\sim$ 15% at 565 ºC down to near zero at 635 ºC. Thus, the latter temperature settles an upper limit for a successful In incorporation for the given amount of active N available for In incorporation, $\Phi_N$ - $\Phi_{Ga}$ - $\Phi_{Al}$. As shown in Fig. 2(b), the In incorporation decrease was accompanied by a decrease of the growth rate. Both effects can be explained by any of the following thermally enhanced phenomena: the dissociation of In-N bonds, or the partial desorption of the impinging In flux (which reduces the amount of In available for QN growth). However, during growth the RHEED pattern was in all cases diffuse and stable indicating the presence of certain In coverage, that is, In was available on the surface for the growth. Hence, the growth rate was not limited by the impinging in flux. This conclusion is supported by the RBS/C data which show, as previously commented, a higher In concentration near the surface in those samples grown at higher temperatures. Therefore, present results point to the dissociation of In-N bonds as the main driving mechanism for the In incorporation decrease when increasing the growth temperature. This conclusion is in good agreement with the results reported by both Böttcher *et al.* and Averbeck and Riechert for InGaN alloys grown by PA-MBE [24,25].

## 2. Effect of the AlN mole fraction

A second set of samples was grown to investigate the effect of the AlN mole fraction on In incorporation (series B). Four different QN samples were grown under intermediate metal-rich conditions at 600 ºC with AlN mole fractions in the 1% to 27 % range. To vary the AlN mole fraction, the impinging In and N fluxes were held constant while the Ga flux was partially replaced by Al, always keeping constant the sum of the Al and Ga fluxes with $(\Phi_{Ga} + \Phi_{Al}) / \Phi_N \sim 75$ %. Thus, the active N available for In incoporation was the same for all samples ($\sim 25$ %). As for series A, the RHEED patterns during growth were streaky and In droplets were not observed on the surface by optical microscopy when finishing growth. Figure 3 shows both the InN mole fraction and the growth rate as a function of the AlN mole fraction, as determined by RBS. Indium incorporation was found to decrease for increasing AlN mole fractions, despite being constant the amount of active N available for In incorporation. As shown in figure 3(b), the decrease of the InN mole fraction was accompanied by a reduction of the growth rate. Because the metal fluxes were adjusted to ensure intermediate metal-rich conditions, the decrease of the growth rate, together with the In incorporation reduction for higher AlN mole fractions, reveals an Al-dependent limit for In incorporation. Present results are in close agreement with those published by Monroy *et al.* (Refs. [8,9]) who claimed that increasing AlN mole fractions enhances InN segregation due to the much higher Al-N bond energy compared to that of Ga-N and specially In-N. We note that, elastic strain may also influence In segregation [8,26]. However, if elastic strain were the dominant force we would expect an inhomogeneous In incorporation along the growth direction due to the progressive relaxation of the growing layer. Such a dependence of the In incorporation on the layer thickness was not observed for any sample from series A or B. Therefore, we suggest that in the studied samples elastic strain plays an inferior role in In segregation.

**B. Structural characterization**

The crystal quality of the QN layers was assessed by HRXRD scans around the (0002) Bragg reflection and by rocking curve scans. No peaks were found in the $\theta/2\theta$ scans except those corresponding to the sapphire substrate, the GaN buffer and the QN layer (no phase separation observed). The full width at half maximum (FWHM) values of the rocking curves around the (0002) Bragg reflection, sensitive to crystal tilt and grain size [8], were analysed as a function of the growth temperature and alloy composition. As shown in figure 4(a), FWHM values steadily decrease with increasing growth temperatures from ~ 435 arcsec @ 565 ºC down to ~ 250 arcsec @ 635 ºC. The latter value is even lower than that of the GaN template, (~ 340 arcsec) revealing that for high enough growth temperatures the crystal mosaicity is at least as good as that of the GaN template. This trend was also confirmed by RBS/C measurements, where minimum yield ($\chi_{min}$) was determined for In and Ga signals close to the surface. The sample grown at 565 ºC showed 34.1 ± 0.3 % of minimum yield and a significant dechanneling, both evidencing defects inside the crystal lattice. Nevertheless, for samples grown above 565 ºC, the minimum yield was drastically diminished, reaching a value of 2.17 ± 0.07 % at 635 ºC, which confirms a very good epitaxial growth of the wurtzite QN.

To analyse the effect of alloy composition on crystal quality, the values of the FWHM were plotted as a function on the AlN/InN mole fraction ratio for a set of $In_xAl_yGa_{1-x-y}N$ samples grown at 600 ºC (similar to series B) with $0.01 < x < 0.08$ and $0.01 < y < 0.27$ [figure 4(b)]. The FWHM values were found to lie between 280 and 365 arcsec with no correlation with alloy composition. RBS/C measurements also demonstrated an almost constant value of the minimum yield ~ 2.3 %, revealing that the quality of the layer is preserved with depth. It is interesting to notice that the sample with the AlN/InN ratio closest to achieve in-plane lattice match to GaN (AlN/InN ~ 4.7 according to Vegard's law) showed the highest FWHM value. Therefore, we conclude that within the composition range studied, the mosaicity of the QN layers is mainly determined by the growth temperature. Present results are opposite to those reported by Wang et al. in Ref. [11] who observed an improvement of the crystal quality in QN samples with higher Al contents (34 % - 45 %) as the AlN/InN mole fraction ratio brought the layers closer to the lattice-match condition.

The crystal quality and the possible presence of alloy fluctuations were also investigated by TEM for series B. Cross-section TEM images (figure 5(a)) revealed that the threading dislocations (TDs) in the QN layers are not generated at the $In_xAl_yGa_{1-x-y}N$/GaN interface. They arise from the GaN template and then extend through the MBE GaN and QN layer until they reach the free surface. In addition, we found no evidences of phase separation (i.e. distint InN, GaN and AlN layers) or In-clustering in the QN layers, as can be observed in figure 5(b), representing a $\mathbf{g}_{0002}$ dark-field TEM micrograph of the sample with 5% In and 10% Al. However, the contrast inside the QN layer was not completely homogeneous (for comparison, see the contrast inside the MBE GaN layer), instead it exhibited a slight "tweed" morphology. This "tweed" pattern was more clearly observed in the sample containing a higher AlN mole fraction (2% In, 27% Al, figure 5(c)) and suggests the presence of alloy fluctuations. The period of such composition fluctuations is in the range of a few nm, as revealed from intensity linescans (figure 5(d)) and depends on the alloy composition: as higher the Al content is, smaller the period of the fluctuations is.

## C. Optical characterization

Figures 6(a) and (b) show the RT-PL spectra for series A and B respectively. A broad emission peak (100 meV < FWHM < 170 meV) from the QN layers was observed for both series. In series A, the emission peak blue-shifted from 2.90 to 3.55 eV due to the band-gap increase caused by a lower In incorporation for increasing growth temperatures. The shift was accompanied by an enhancement of the integrated PL intensity, pointing to a steadily reduction of the density of NRCs. This assumption is supported by the improvement in crystal quality observed by both HRXRD and RBS/C. For series B, the emission energy shifted from 3.20 to 3.45 eV for increasing AlN mole fractions and the integrated PL intensity showed a considerable increase. Dependent PL measurements in similar samples, recently published in Ref. [27], showed that the emission energy and band-width followed S- and W-shaped temperature dependences respectively, typical of exciton localization by alloy inhomogeneities. For a given InN mole fraction, such shapes were found to shift towards higher temperatures with increasing AlN mole fractions, indicating a stronger exciton localization (deeper) for higher Al contents. This fact also provides an explanation to the observed RT-PL enhancement with increasing AlN mole fractions, since excitons with a stronger localization are less defect-sensitive [3,5]. The presence of alloy fluctuations and their dependence with the AlN mole fraction were corroborated by TEM as previously discussed.

Although it is well established that exciton localization in In containing III-nitrides takes place at "In-rich regions" within the In(Al,Ga)N matrix, the nature of such regions remains under debate and may depend on the specific growth technique and conditions [5]. CL measurements were carried out to investigate the origin of exciton localization in these QN layers. As a representative example, figure 7 summarizes the results from CL characterization at 5 K using an accelerating voltage of 10 kV and a spot diameter of ~1 μm for a QN sample grown at 600 °C with InN and AlN mole fractions of 3 % and 19 % respectively. Figures 7(a) and (b) show CL emission wavelength and intensity mappings over an area of 15 μm x 25 μm. Small variations were observed in both wavelength and intensity, in contrast with results by Cremades et. al. in Ref. [28] who observed a microscopic island-like emission from QN samples grown by PA-MBE at much higher temperatures (650 - 775 °C). In figure 7(c), the spatially averaged spectrum (over the 15 μm x 25 μm area) and a histogram depicting the frequency of the CL emission peak wavelengths shown in figure 7(a) are displayed. Two peaks were observed at 3.5 and 3.6 eV, corresponding to the GaN template and the QN layer respectively. To statistically analyze the fluctuations of the QN layer quantitatively we computed the histogram of the wavelength image: The frequentness of an energy is plotted versus this photon energy for all 51,200 pixels. The depicted histogram shows a Gaussian-like function (random-statistical distribution) over 4 orders of magnitude with a standard deviation of $\sigma(E_{CL})$ = 150 meV. The macroscopic line shape is a direct summation of many sharper nanoscopic emission lines. Thus, the macroscopic intensity profile directly represents the probability distribution of the nanoscopic emissions. To elucidate the origin of the CL broadening, a CL local spectrum [also shown in fig. 7 (c)] was acquired with much higher spatial resolution using a spot diameter below 150 nm. The local spectrum was found almost identical to the macroscopic integral spectrum, revealing that the nanoscopic fluctuations responsible of the CL broadening must have a lateral length scale smaller than the spatial resolution of the CL experiment (~150 nm). Otherwise, if fluctuations were spatially resolved, the local CL spectrum should be narrower than the integral CL one. Again, the CL results are in good agreement with the TEM measurements where composition fluctuations in the range of 10-15 nm were detected.

## IV. SUMMARY AND CONCLUSIONS

In summary, wurtzite (0001)-oriented $In_xAl_yGa_{1-x-y}N$ layers with $0.01 < x < 0.15$ and $0.01 < y < 0.27$ were grown by PA-MBE on GaN templates grown on $c$-plane sapphire. RBS measurements showed that despite segregation, In was homogeneously incorporated along the growth direction. Moreover, channeling measurements demonstrated that good crystal quality was preserved whithin the layer. The maximum amount of In that can be incorporated was found to be limited by both the growth temperature, through thermally enhanced In-N bonds dissociation, and the AlN mole fraction. The FWHM values of the rocking curves of the (0002) symmetric reflection highlighted that the crystal mosaicity of the QN layers is mainly determined by the growth temperature, being independent of the alloy composition within the composition range studied. No evidences of phase separation were observed, either by HRXRD or TEM, although there are evidences of nanometer-sized composition fluctuations highly dependent on the Al content. PL spectra from the QN layers at RT were found to be dominated by a broad emission line. The emission peak energy and bandwidth showed respectively S- and W-shapes temperature dependences, typical of exciton localization by alloy inhomogeneities. CL results demonstrated that such alloy inhomogeneities within the InAlGaN matrix take place on a lateral length scale below 150 nm, in agreement with TEM investigations.

**ACKNOWLEDGMENTS**

Thanks are due to L. Cerruti, M. Utrera, F. González-Posada, J. Grandal, and A. Bengoechea for fruitful discussions. This work was partially supported by research grants from the Spanish Ministry of Education (NAN04/09109/C04/2, Consolider-CSD 2006-19, and FPU program); and the Community of Madrid (S-0505/ESP-0200 and Futursen).

**List of Figures**

presence of alloy fluctuations. (d) Intensity line-scan on the area marked in (c) revealing that the period of the alloy fluctuations is in the range of a few nm.

Figure 6    (a) RT photoluminescence spectra for QN layers grown at different temperatures in the 565-635 ºC range. (b) RT photoluminescence spectra for a set of QN samples grown with increasing AlN mole fractions at 600 ºC.

Figure 7    Cathodo-luminescence emission wavelength (a) and intensity (b) maps acquired over an area of 15 μm x 25 μm. (c) Integral and local CL spectra for the QN layer and the GaN template together with the histogram generated from the frequency of the CL emission peak wavelengths shown in (a).

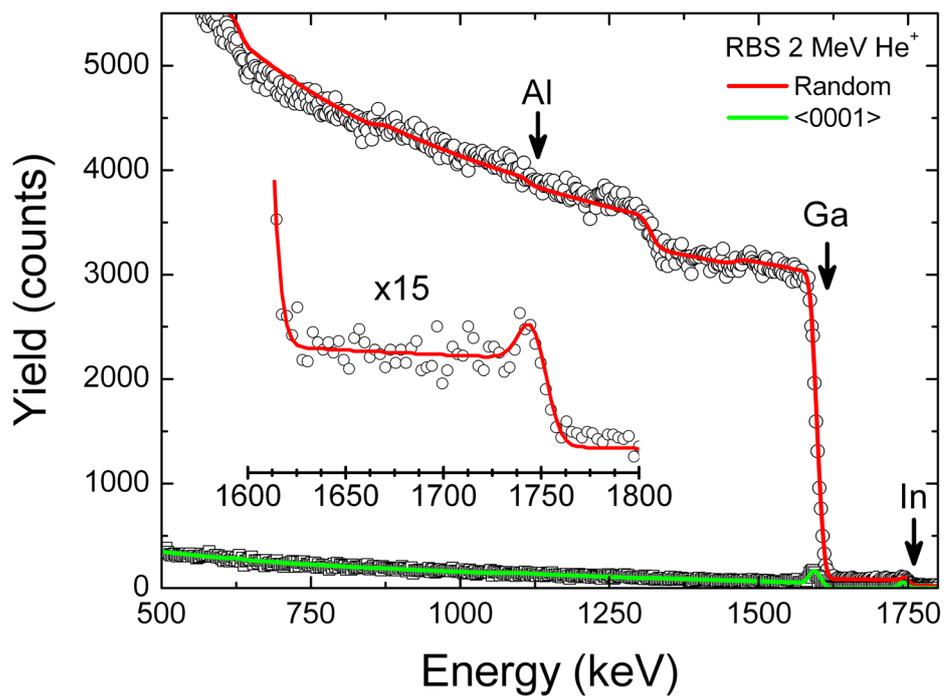

**Figure 1**

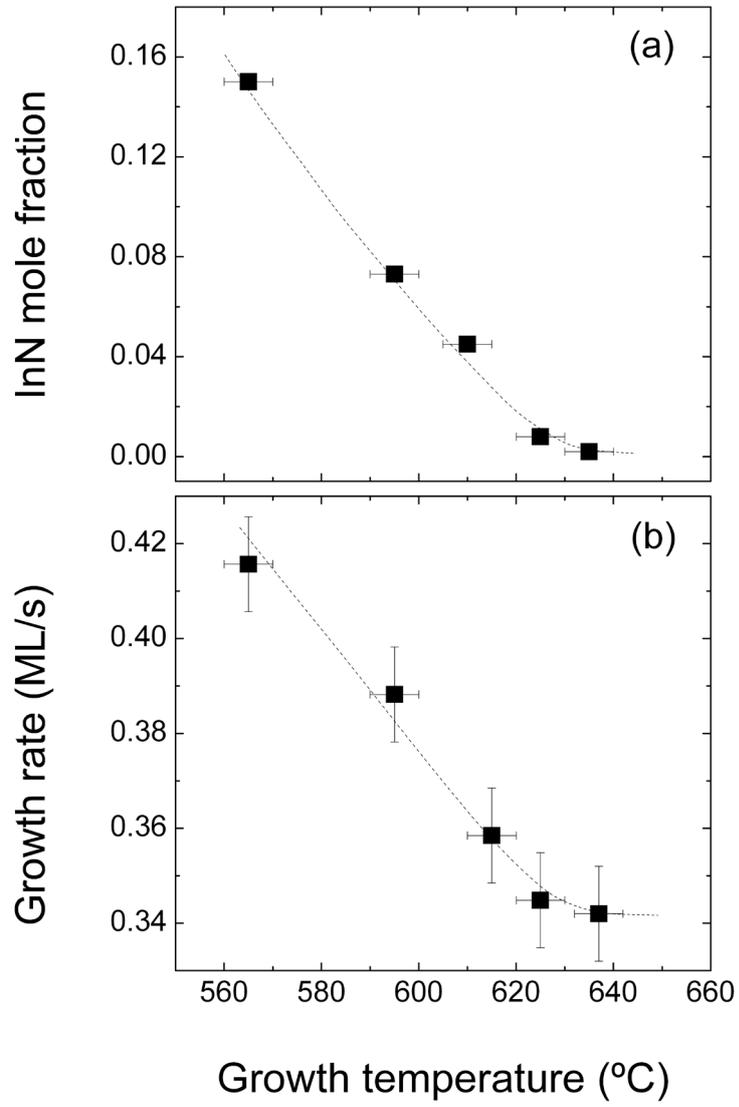

**Figure 2**

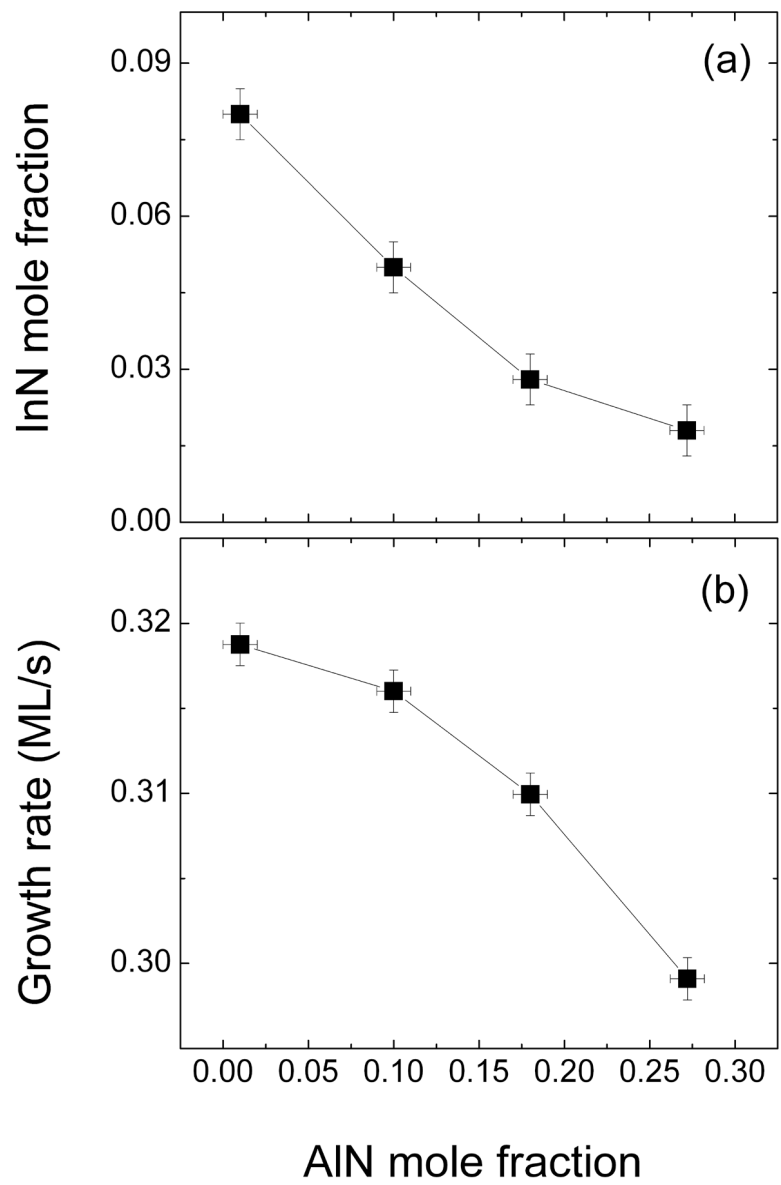

Figure 3

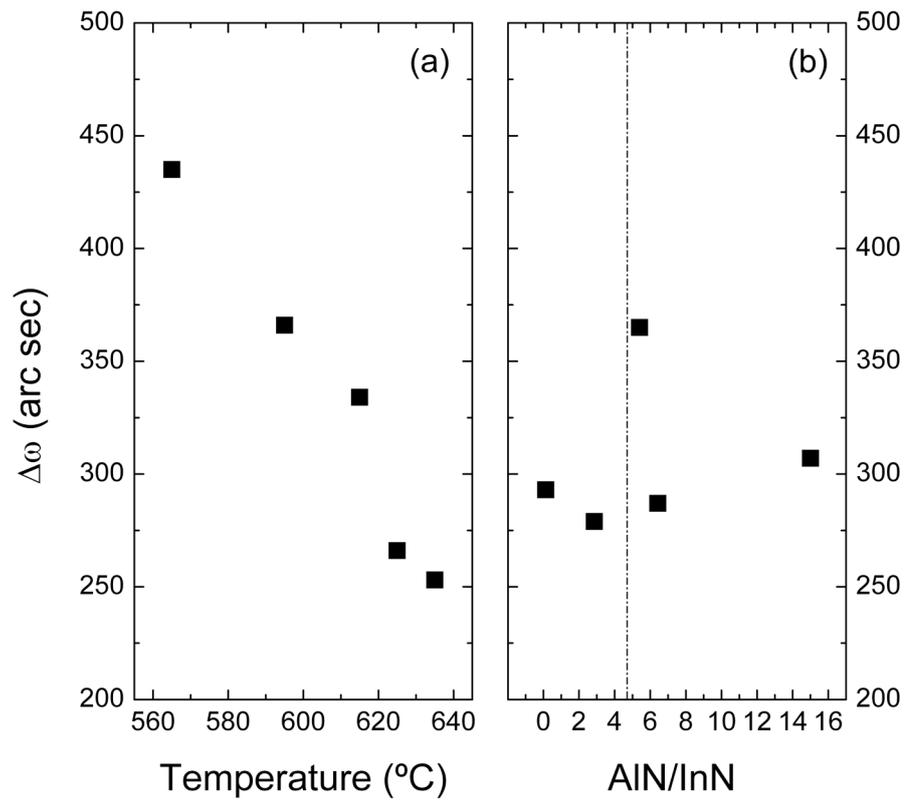

**Figure 4**

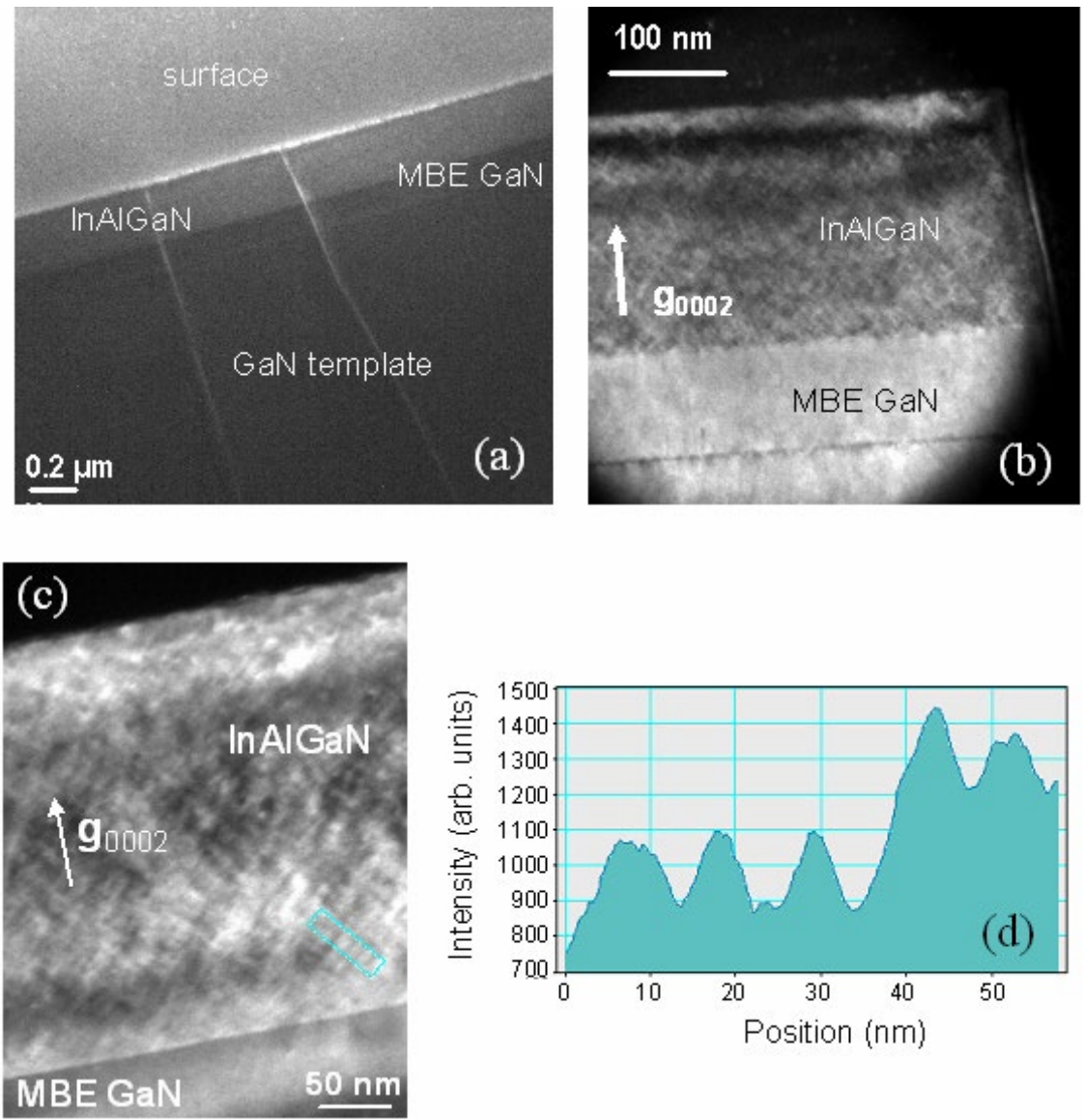

**Figure 5**

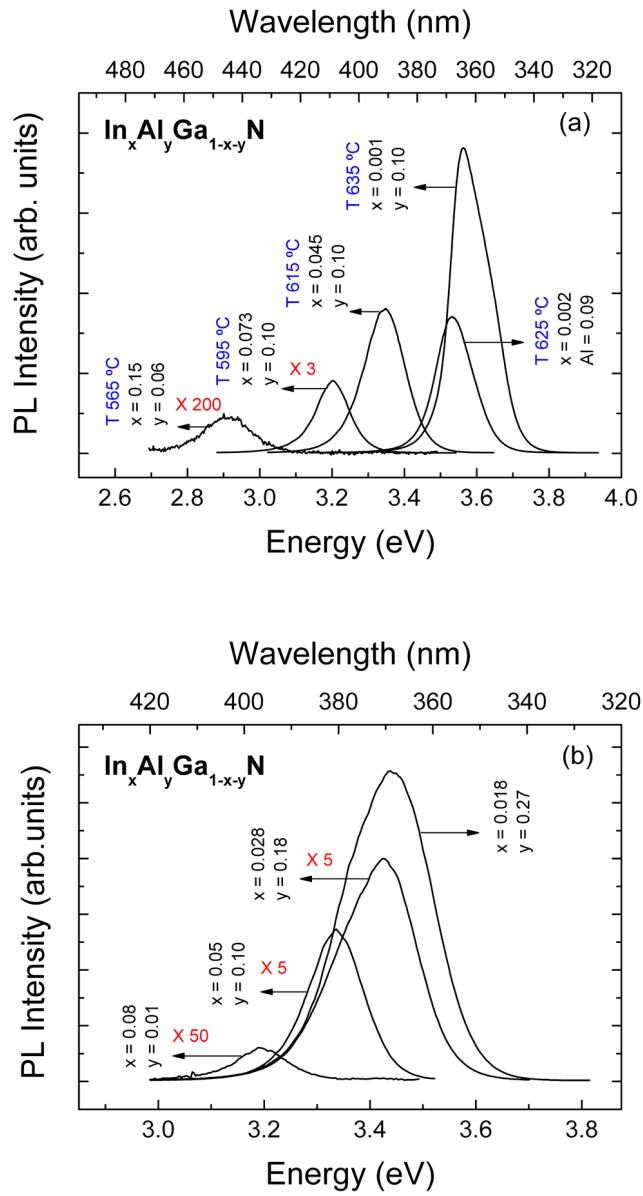